# Carrier-Dopant Exchange Interactions in Mn-doped PbS Colloidal Quantum Dots

Gen Long, Biplob Barman, Savas Delikanli, Yu Tsung Tsai, Peihong Zhang, Athos Petrou, and Hao Zeng*

Department of Physics, SUNY at Buffalo, Amherst, NY, 14260

**Abstract**

Carrier-dopant exchange interactions in Mn-doped PbS colloidal quantum dots were studied by circularly polarized magneto-photoluminescence. Mn substitutional doping leads to paramagnetic behavior down to 5 K. While undoped quantum dots show negative circular polarization, Mn doping changes its sign to positive. A circular polarization value of 40% was achieved at T=7 K and B=7 tesla. The results are interpreted in terms of Zeeman splitting of the band edge states in the presence of carrier-dopant exchange interactions that are qualitatively different from the *s,p-d* exchange interactions in II-VI systems.

*Author to whom correspondence should be addressed. Electronic mail: haozeng@buffalo.edu.

In diluted magnetic semiconductors (DMS), carrier-dopant exchange interactions lead to carrier spin polarization, which can be harnessed for spintronics applications.[1] In quantum dots (QDs), confinement of carrier wave functions can profoundly modify such interactions, leading to exotic properties.[2] Colloidal QDs are of particular interest since they are free of a matrix material, provide the strongest confinement potential, and are convenient for engineering the carrier and dopant wave function overlap in core/shell heterostructures. Recently, spin properties of colloidal QDs have received significant attention. For example, Beaulac *et al.* reported light induced spontaneous magnetization up to room temperature in Mn-doped CdSe QDs;[3] Bussian *et al.* reported tunable magnetic exchange interactions in Mn-doped inverted core–shell ZnSe–CdSe nanocrystals.[4] The effective exciton $g$-factors can be tuned in both magnitude and sign, due to the interplay between quantum confinement and wave function engineering. Both References 2 and 3 studied Mn-doped II-VI systems, where the well-established *s,p-d* exchange mechanism can explain the magnetooptical results.

Lead salts such as PbS belong to IV-VI narrow gap semiconductors. They have unique electronic, optical and thermal properties with potential applications in photovoltaic, optoelectronic and thermoelectric devices.[5-7] Different from the extensively studied II-VI systems where the band edges are located at the $\Gamma$ point, the band edge states of lead salts with a rock salt structure are four-fold degenerate (eight-fold degenerate including spin) at the *L* point of the Brillouin zone.[8,9] Many of their interesting properties are associated with their small band gaps and strong spin-orbit interactions. The strong relativistic effects substantially contract the wave function of the Pb *6S* states, pushing it into the valence band, and thus the conduction band minimum is derived mostly from the Pb *6p* states.[10] This changes the carrier-dopant exchange interactions in transition metal doped lead salts qualitatively. In addition, in QDs, quantum confinement tuning of the wave function and carrier-dopant interactions would lead to interesting magnetooptical properties. Surprisingly, while undoped IV-VI QDs have been extensively studied for their multiple exciton generation properties[11] and applications in photodetectors[12] and photovoltaics,[13] magnetooptical studies of transition metal doped IV-VI QDs are scarce in literatures.

In this paper, we report the carrier spin polarization of Mn-doped PbS colloidal QDs measured by circularly polarized magneto-photoluminescence (CP-MPL). A maximum circular polarization ($P$) value of 40% has been obtained at T=7 K and B= 7 tesla. The sign of $P$ is found to be positive, which is opposite to the sign of $P$ from undoped PbS. This is attributed to the carrier-dopant exchange interactions in this system, which leads to a decrease in the effective electron $g$-factor while an increase in the hole $g$-factor, thus inverting the energy of the σ+ and σ- transitions.

$Pb_{1-x}Mn_xS$ (x=0.005-0.03) QDs were synthesized using the organic solution phase technique, by modifying the published procedure for undoped PbS QDs.[14] The detailed synthesis procedure will be published elsewhere. As-synthesized QDs have the rock-salt structure, as identified by the X-ray diffraction (XRD) pattern shown in Fig. 1(a). Transmission electron microscope (TEM) images (Fig. 1(b)) shows that these QDs are essentially spherical, with sizes tunable from 2.5 nm to 10 nm. High resolution TEM images (Fig. 1(c)) show lattice fringes with spacings of 3.4 Å and 3.0 Å, corresponding to (111) and (200) planes of the rock-salt structure, respectively. The composition of the particles was characterized by Energy dispersive X-ray spectroscopy after purifying the nanoparticles and removing the surface bound $Mn^{2+}$ ions with pyridine. The maximum $Mn^{2+}$ doping concentration is around 3%. Two samples: 4 nm undoped PbS QDs and 4 nm $Pb_{0.97}Mn_{0.03}S$ QDs were further investigated using magnetometry and magneto-photoluminescence spectroscopy.

Undoped PbS QDs are diamagnetic, as measured by the VSM option in a Quantum Design PPMS system. In contrast, Mn-doped QDs give paramagnetic signals at temperatures down to 5 K (Fig. 2), indicating successful Mn incorporation. The 5 K magnetization at 9 T is about 0.55 emu/g, comparable to results in previous reports.[15] This corresponds to the Mn moment of 4.2±0.4 $\mu_B$. This value is lower than 5 $\mu_B$ due to the fact that the magnetization is not saturated at 9 tesla and that there exists effective antiferromagnetic coupling between $Mn^{2+}$ ions. A modified Brillouin function, as shown in Eq. (1) is used to fit the magnetization curves measured at different temperatures[16]

$$M = xN_0 g\mu_B SB_S[\frac{g\mu_B SB}{k_B(T+T_o)}] \tag{1},$$

where $x$ is the doping concentration, $N_0$ is the number of cations per unit cell, $\mu_B$ is the Bohr magneton, $S = 5/2$ is the spin of $Mn^{2+}$ and $B_S$ is the Brillouin function, $B$ is the applied magnetic field, $T$ is the lattice temperature and $T_o$ is the effective temperature taking into consideration of the $Mn^{2+}$- $Mn^{2+}$ interactions. The fitted effective temperature $T_o \sim +3.0$ K, which is consistent with the antiferromagnetic coupling between $Mn^{2+}$ ions. Note here that the magnetization curve is different from that with photoexcitation, since in the presence of photo-generated carriers, magneto-polaron is formed, which can generate a large internal field.[17]

Photoluminescence and CP-MPL were measured in a 7 tesla variable temperature optical cryostat. The PL was excited at 785 nm at a laser power of 1 mW. The emitted light was collected in the Faraday geometry and focused onto the entrance slit of a single monochromator equipped with an InGaAs multichannel detector. The light was analyzed in its left ($\sigma$+) and right ($\sigma$-) circularly polarized components by a combination of quarter wave filter and linear polarizer placed just before the entrance slit. The circular polarization P is defined as $\frac{I_{\sigma+}-I_{\sigma-}}{I_{\sigma+}+I_{\sigma-}}$, where $I_{\sigma+}$ and $I_{\sigma-}$ are the intensity of left and right circularly polarized emission, respectively. For both undoped and doped QDs with sizes of 4 nm, the emission is centered at 950 meV and has a full-width-at-half-maximum of 90 meV at 7 K. The $\sigma_+$ and $\sigma_-$ components of the PL spectra from Mn-doped PbS QDs at zero magnetic field are shown in Fig. 3(a). The two circularly polarized components have equal intensity and thus give a circular polarization $P = 0$. When a magnetic field is applied the intensity of the $\sigma_+$ component becomes higher than that of the $\sigma_-$, as is shown in Fig. 3(b), resulting in a positive circular polarization $P = 40\%$ at B=7 tesla.

For undoped PbS QDs, the field dependence of circular polarization is linear and shows a negative slope, as shown in Fig. 3(c). $P = -25\%$ at $T = 7$ K and $B = 7$ tesla. The Zeeman splitting of the band edge states is shown in Fig. 4(a), following an earlier work on PbSe QDs.[9] The sign of the polarization P can be determined by the exciton g-factor $g_{ex} = g_e - g_h$. With $g_e$ and $g_h$ both being positive and $g_e > g_h$, $g_{ex}$ is positive, and the $\sigma$- transition has lower energy than that of $\sigma$+, leading to negative P. Using the simple Brillouin function $P = -\tanh(\frac{g_{ex}\mu_B B}{2k_B T}) \approx -\frac{g_{ex}\mu_B B}{2k_B T}$, since P vs. B is linear, $g_{ex}$ is obtained to be $+0.77\pm0.05$.[9] The measured polarization P and fitted $g_{ex}$ are twice as large as those reported in the earlier reference.[18] The discrepancy can probably be attributed to a difference in particle size. It is expected that $g_{ex}$ should decrease with decreasing particle size, due to quantum confinement induced increase in the band gap and a reduction in the spin-orbit coupling. However, the experimentally reported size dependence of $g_{ex}$ in PbSe[19] and PbS[18] QDs show opposite trends. It is not clear what physics leads to an increase in $g_{ex}$ for PbS QDs with decreasing particle size reported in Ref. 18.

With Mn doping, the $P$ values turn positive, i.e. the intensity of $\sigma+$ transition is now higher than that of $\sigma-$, as shown in Fig. 3(c). $P$ as a function of magnetic field $H$ measured at different temperatures can be fitted by Eq. (2)

$$P = -P_o \tanh(\frac{\Delta E}{2k_B T}) \qquad (2),$$

where $P_o$ is a pre-factor that takes into account the random particle orientation and non-uniform doping. The excitonic Zeeman splitting $\Delta E$ can be given by Eq. (3)

$$\Delta E = g_{ex}\mu_B B + xN_0(\Delta\varepsilon_{VBM} - \Delta\varepsilon_{CBM})SB_S(\frac{g\mu_B SB}{k_B T}) \qquad (3),$$

where the first term represents the intrinsic Zeeman splitting and the second term describes the splitting due to the carrier-dopant exchange interactions, which is proportional to the magnetization. $\Delta\varepsilon_{VBM}$ and $\Delta\varepsilon_{CBM}$ are the exchange splitting energy of the VBM and CBM, respectively. For simplicity, only band edge states corresponding to $\sigma+$ and $\sigma-$ emissions are considered. Note that we cannot use $(T+T_o)$ in the argument of the Brillouin function in Eq. (3), since carrier-$Mn^{2+}$ interactions compete with $Mn^{2+}$ direct exchange in determining the magnetization. Both effects are expected to be small for low doping concentrations, and are neglected in Eq. (3). The fitted ($\Delta\varepsilon_{VBM}-\Delta\varepsilon_{CBM}$) is 6 meV. At $T$=7 K and $B$= 7 T, this corresponds to $\Delta E$ of 1.8 meV, which is beyond the energy resolution of our instrument. This is why we do not see a clear energy separation in the $\sigma+$ and $\sigma-$ emissions (Fig. 3(b)). To compare with the giant Zeeman splitting observed in CdMnSe systems,[1] we extracted the effective $g$-factors using the linear portion of the fitting curve for $P$ vs. $B$. The $g_{eff}$ values are estimated to be -6.0, -4.3 and -1.7 at 7, 14 and 25 K, respectively. The temperature dependence of $g_{eff}$ is consistent with earlier reports in Mn-doped lead salts ($g_{eff}$ can be obtained from the difference in $g_e$ and $g_h$ as a function of temperature).[16] A monotonic decrease in $g_{eff}$ with increasing $Mn^{2+}$ concentration and a change in its sign at a critical concentration are expected;[20] this will be investigated in the future.

A change in the sign of polarization $P$ represents an inversion of the energy for $\sigma+$ and $\sigma-$ transitions. Below we discuss the mechanism responsible for the observed inversion. In Mn-doped II-VI systems, a sign change in $P$ value is also observed.[2] There the situation is quite clear. The Zeeman splitting of the band edge electron ($J=1/2$) and heavy hole states ($J=3/2$) of undoped CdSe is shown in Fig. 4(c).[2] $g_{ex} = (-3g_{hh}-g_e)$ ($g_{hh}$ is negative by definition) is positive, which means that $\sigma+$ has higher energy than $\sigma-$ and $P$ is negative. In Mn-doped CdSe, the carrier-dopant exchange interactions can be described by two parameters in a mean field model: the *s-d* coupling parameter $\alpha$ and *p-d* coupling parameter $\beta$.[1] $\alpha$ is positive (ferromagnetic) and $\beta$ is negative (antiferromagnetic), while $\beta \cong -4\alpha$. With increasing doping concentration, the exchange splitting will dominate the intrinsic Zeeman splitting. At high enough doping concentration, the Zeeman splitting of the conduction band minimum (CBM) is slightly increased while that of the valence band maximum (VBM) is reversed, as can be seen from Fig. 4(d). Because *p-d* coupling is the dominant term, the $\sigma+$ transition will now have lower energy than the $\sigma-$ transition, and therefore $P$ becomes positive.

The situation in IV-VI systems is more complicated. The band gap is at the *L* point, which has multi-valley degeneracy. The carrier-dopant exchange interactions can be described by the well-known $\vec{k} \cdot \vec{p}$ model, which needs four exchange parameters: two for the holes and two for the electrons.[21] Depending on field orientations, the electron and hole g-factors are anisotropic for different valleys. Since our QDs are randomly oriented, the circular polarization should be determined by the effective *g*-factors averaged over all directions. The inter-band MPL cannot determine $g_e$ and $g_h$ independently, we therefore use first principles techniques to evaluate the exchange splitting. According to our calculation, the effective *p-d* interactions between Mn and VBM is antiferromagnetic in nature, i.e., the energy of the majority spin VBM state is higher than that of the minority spin VBM state, and the splitting is of the order of 100 meV. This can be understood since virtual hopping can only occur between the minority spin VBM and unoccupied Mn *d* states. This will lead to an increase in the Zeeman splitting of VBM with Mn doping. On the other hand, the exchange between Mn and the CBM states is ferromagnetic. It can be understood in terms of an enhanced exchange potential for the majority spin state. This means that Mn doping leads to an increase in the Zeeman splitting of CBM as well. The exchange splitting is weak, of the order of 10 meV, since the CBM states are derived from Pb *p* states which overlap only slightly with Mn *d* states. It is thus expected that the carrier-dopant exchange interactions are dominated by *p-d* interactions between Mn and valence band electrons. According to our calculations, we can construct the schematic diagram of the Zeeman splitting of the band edge states for Mn-doped PbS. As can be seen from Fig. 4(b), both the effective $g_e$ and $g_h$ increase with increasing doping concentration; however, the enhancement in $g_h$ dominates. At high enough doping concentration, the carrier-dopant exchange interactions should lead to $g_h > g_e$. This causes the polarization *P* to reverse its sign from negative to positive. We notice that in the earlier experimental work, Mn doping leads to a decrease in $g_e$, which is opposite to our theoretical predictions.[16] Both scenarios will result in positive *P*. However, we argue that it is more plausible that both $g_e$ and $g_h$ increases with increasing Mn doping (Fig. 4(b)), and therefore the exchange splittings of CBM and VBM oppose each other. This is based on the fact that our observed $|g_{eff}|$ is of the order of 10, which is an order of magnitude smaller than that in doped II-VI systems. In Mn-doped CdSe, the exchange splittings of CBM and VBM both enhances the excitonic splitting (Fig. 4(d)), leading to a giant Zeeman splitting with $g_{eff}$ of the order of 100.

In conclusion, we observed positive circular polarization of about 40% in Mn-doped PbS colloidal QDs. The sign of the polarization is opposite to that of the undoped system which shows a polarization value of -25% at the same temperature and field. The sign change is attributed to the enhancement in both the valence and conduction electron $g$ factors due to carrier-dopant exchange interactions, dominated by $g_h$ enhancement. While both IV-VI and II-VI systems show an inversion of the circular polarization upon Mn doping, the involved band edge states and Zeeman splitting of these levels are qualitatively different. Future work will be devoted to a systematic study of the role of quantum confinement on tuning the exchange interactions.

**Acknowledgement**

The authors would like to thank Dr. Yuelin Qin for TEM measurements and Rafal Oszwaldowski for helpful discussions. This work is supported in part by NSF DMR-0547036 and Department of Energy under Grant No. DE-SC0002623.

**Figure captions:**

Figure.1 (a) A typical XRD pattern of Mn-doped PbS QDs, showing the rock-salt structure with sizes of 4 nm and doping concentration of x=0.02; (b) A TEM image of typical Mn-doped PbS QDs; and (c) a HRTEM image of Mn-doped PbS QDs, showing the (111) and (200) lattice friges, with lattice spacings of 3.5 and 3.0 Å, respectively.

Figure. 2 Magnetization as a function of magnetic field for Mn-doped PbS QDs measured at 5, 8, 10, 50, 100 and 293 K, respectively.

Figure. 3 Circularly polarized photoluminescence spectra for Mn-doped PbS QDs measured at T = 7 K and (a) B = 0; (b) B = 7 tesla; (c) Circular polarization *P* as a function of magnetic field for Mn-doped PbS QDs measured 7, 14, and 25 K, respectively and for undoepd PbS QDs measured at 7 K; the lines are fitting curves using Eq.(2).

Figure. 4 Schematic band diagram and inter-band transitions in the presence of a magnetic field for (a) undoped PbS; (b) Mn-doped PbS; (c) undoped CdSe and (d) Mn-doped CdSe.

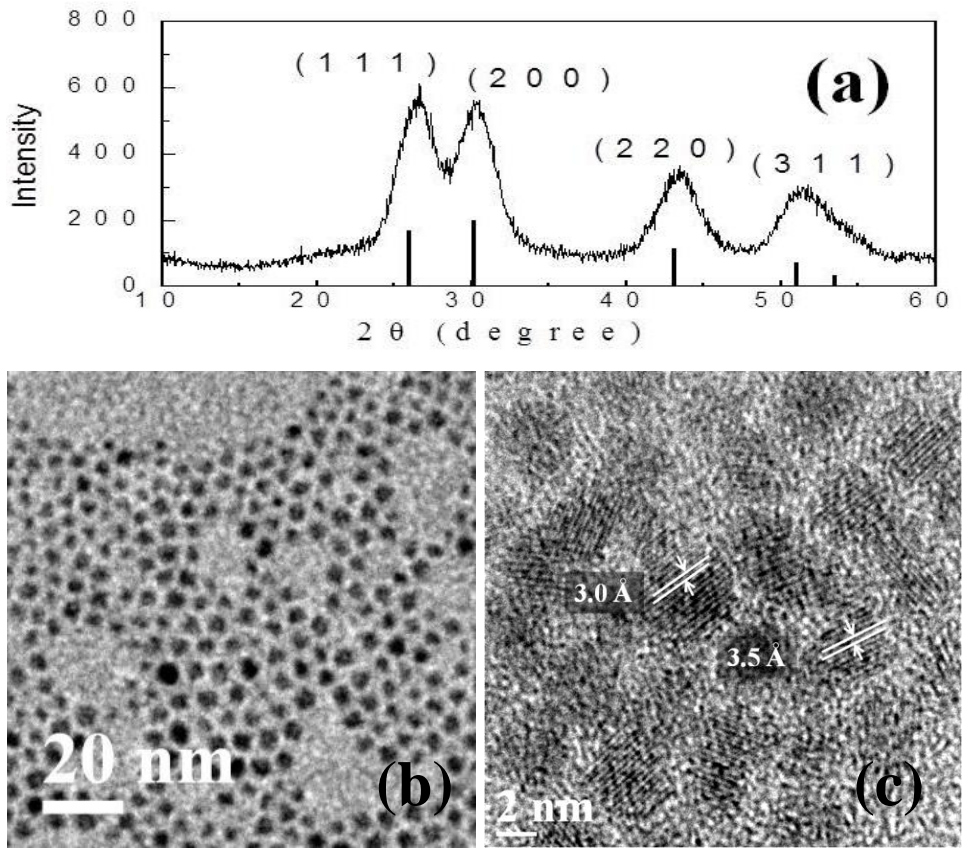

Figure. 1

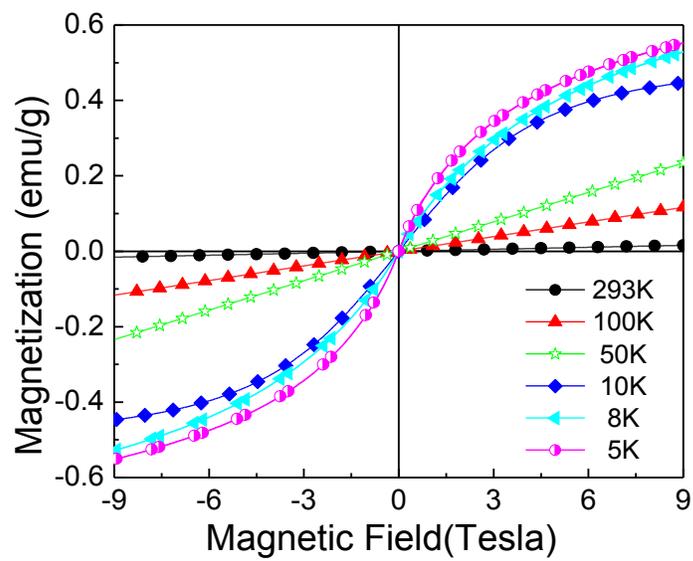

Figure. 2

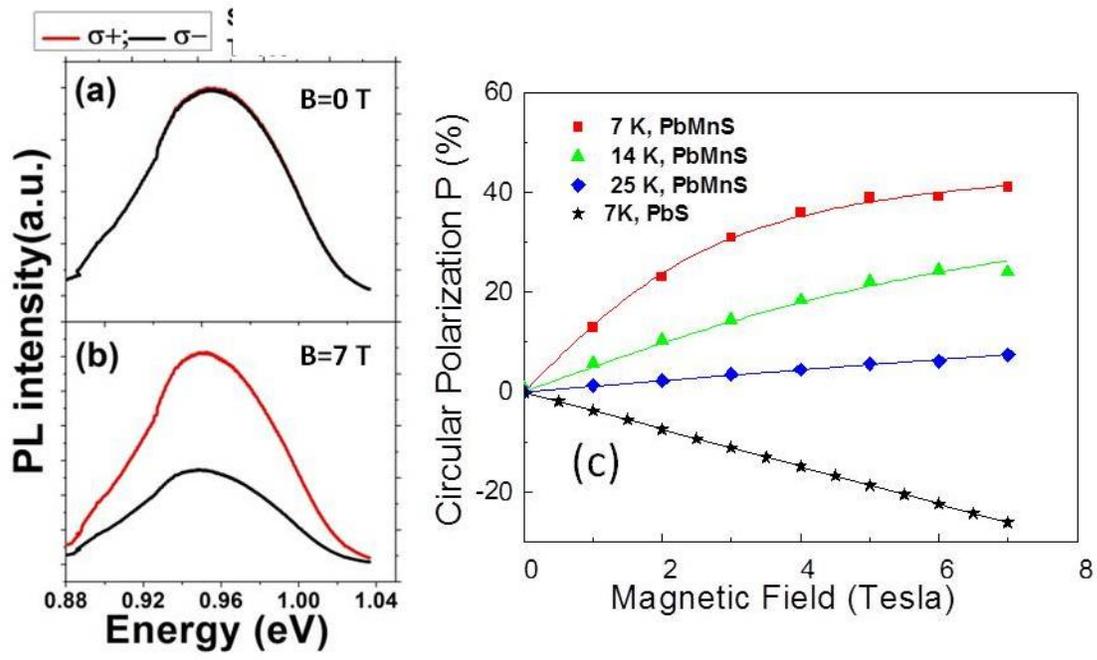

Figure. 3

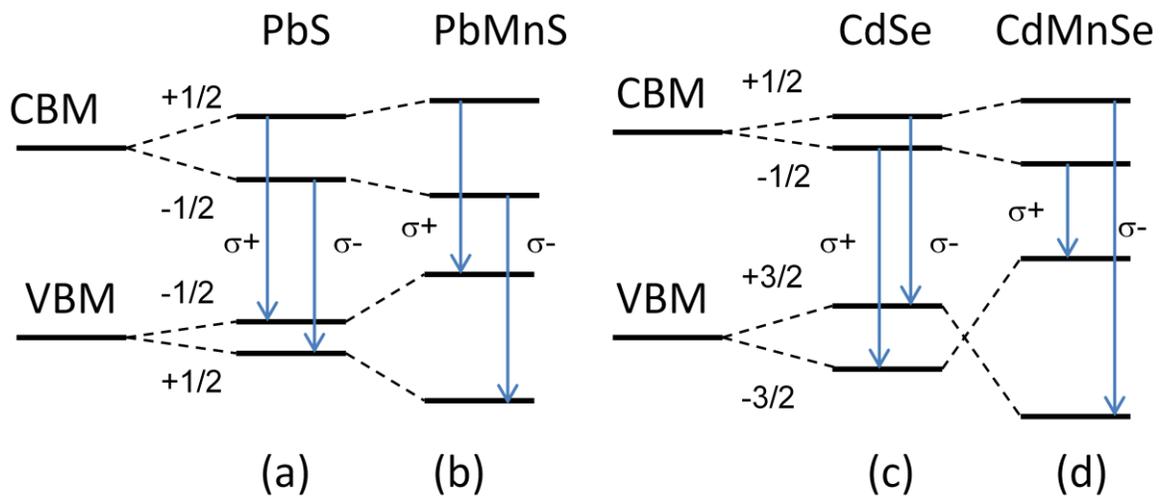

Figure. 4